\documentclass[final,5p,times,twocolumn]{elsarticle}

\usepackage{amssymb}

\usepackage[table,xcdraw]{xcolor}
\usepackage[normalem]{ulem}
\usepackage{array}
\newcolumntype{x}[1]{>{\centering\arraybackslash\hspace{0pt}}p{#1}}

\definecolor{ao}{rgb}{0.0, 0.5, 0.0}

\PassOptionsToPackage{hyphens}{url}
\usepackage[hidelinks]{hyperref}
\hypersetup{
    colorlinks,
    linkcolor=blue,
    citecolor=blue,
    urlcolor=blue
}
\expandafter\def\expandafter\UrlBreaks\expandafter{\UrlBreaks
    \do\a\do\b\do\c\do\d\do\e\do\f\do\g\do\h\do\i\do\j%
    \do\k\do\l\do\m\do\n\do\o\do\p\do\q\do\r\do\s\do\t%
    \do\u\do\v\do\w\do\x\do\y\do\z\do\A\do\B\do\C\do\D%
    \do\E\do\F\do\G\do\H\do\I\do\J\do\K\do\L\do\M\do\N%
    \do\O\do\P\do\Q\do\R\do\S\do\T\do\U\do\V\do\W\do\X%
    \do\Y\do\Z\do\*\do\-\do\~\do\'\do\"\do\-}%
\usepackage[hyphenbreaks]{breakurl}

\journal{}

\begin{document}

\begin{frontmatter}



\title{Evaluating tamper resistance of digital forensic artifacts during event reconstruction}

\author[1]{Céline Vanini}
\ead{celine.vanini@unil.ch}

\author[2]{Chris Hargreaves}
\ead{christopher.hargreaves@cs.ox.ac.uk}

\author[1]{Frank Breitinger}
\ead{frank.breitinger@unil.ch}
\ead[url]{https://FBreitinger.de}

\address[1]{School of Criminal Justice, 
  University of Lausanne, 1015 Lausanne, Switzerland}

\address[2]{Department of Computer Science, University of Oxford, Wolfson Building, Parks Road, Oxford OX1 3QD, United Kingdom}

\begin{abstract}
Event reconstruction is a fundamental part of the digital forensic process, helping to answer key questions like who, what, when, and how. A common way of accomplishing that is to use tools to create timelines, which are then analyzed. However, various challenges exist, such as large volumes of data or contamination. While prior research has focused on simplifying timelines, less attention has been given to tampering, i.e., the deliberate manipulation of evidence, which can lead to errors in interpretation.
This article addresses the issue by proposing a framework to assess the tamper resistance of data sources used in event reconstruction. We discuss factors affecting data resilience, introduce a scoring system for evaluation, and illustrate its application with case studies. This work aims to improve the reliability of forensic event reconstruction by considering tamper resistance.

\end{abstract}



\begin{keyword}
Event Reconstruction \sep Resistance \sep Tampering \sep Timeline \sep Digital Traces \sep Terminology \sep Factors 



\end{keyword}

\end{frontmatter}



\section{Introduction}
Event reconstruction is a fundamental phase in digital forensic investigations where examiners attempt to answer the questions of who, what, when, whom/what with, where, and how after a crime or incident occurred \cite{roux_sydney_2022}. The reconstruction process often starts with the creation of a timeline using automatic tools such as \emph{Plaso}\footnote{\url{https://github.com/log2timeline/plaso}}, or other (commercial) tools. These tools extract information contained within the file system as well as application-related files and then chronologically organize the data from these different sources. 

Timeline analysis is the second most commonly used digital forensic technique, after keyword searching \cite{hargreaves2024dfpulse}. However, analyzing timelines poses significant challenges, particularly due to the large amount of information they contain, which makes the process time-consuming. Prior research has often concentrated on methods to reduce timeline complexity, such as filtering, labelling \cite{studiawan_sentiment_2020}, and aggregation \cite{hargreaves_automated_2012}.

Analysis of timelines relies on the extracted timestamps being correct, but these timestamps, like all digital evidence, can be vulnerable to tampering. Despite its importance, the issue of tampering, i.e., the deliberate manipulation of evidence by adversaries, has received less attention. In timelines, this can result in incorrect ordering, aggregation, or filtering of entries, leading to substantial errors in interpretation.
When tampering occurs, the risk of misinterpretation rises significantly \cite{CASEY2020200888}, an error that was not explicitly covered in the consideration of the `timeline analysis' technique in recent work on tool error in \cite{hargreaves2024abstract}.

Tampering is not a fictive problem. For instance, \citet{wada2020cas} discusses the World Anti-Doping Agency vs.~Russian Anti-Doping Agency case. The examiners had to analyze MySQL databases using MyISAM storage engine on an Ubuntu Server and developed an approach to find alteration and/or deletion of database records: 
(1) Recovery (carving) to obtain historical backups of MySQL databases (e.g., MySQL dumps, file-level backups); 
(2)	comparing the recovered database versions to detect specific alterations of records, enabling targeted analysis (rows added, deleted, altered and/or copied in each table); 
(3)	targeted analysis of database tables to create customized content carving methods to recover historical data; 
(4)	in-depth analysis of recovered database table structures to detect specific records only existing in the deleted state, as well as out-of-sequence records due to overwritten records; 
(5)	comparing timestamps in databases and on the file systems to detect backdating. 
While for this case the examiners had only one source (databases), several sources may be encountered during an investigation providing divergent information.

To help investigators interpret data, the C-Scale (also referred to as the `Strength of Evidence scale') may be applied \cite{CASEY2020200888}. The scale aims at helping practitioners to express their evaluative opinion in a more understandable and refined manner, at the final stages of the investigation. It involves two essential components: the number of sources that align and their resistance to tampering. As outlined by the C-Scale, evidence becomes stronger when multiple independent sources agree, particularly if these sources are tamper-proof or more difficult to manipulate.

The C-Scale is a powerful resource but it requires practitioners to differentiate between tamper-resistant and not tamper-resistant sources. 
While some investigators may intuitively consider this, research to date has not attempted to evaluate the tamper resistance of artefacts (data sources) used as the basis for event reconstruction. Therefore, this paper aims to bridge that gap and explores factors that could be used to formally evaluate the resistance against tampering of artefacts used as the basis for event reconstruction. 

In summary, this article provides the following \emph{contributions}: we assess the resilience of artefacts by providing an extensive discussion of factors that may affect their resistance to any active modifications and/or deletions in a contextual manner. Additionally, we propose a scoring system that can be used to support the evaluation (Sec.~\ref{sec:factors}). Ultimately, we illustrate the use of the scoring with a set of case study examples (Sec.~\ref{sec:case_study}).


%



\section{Terminology}
\label{sec:terms/def}
\citet{jaquet-chiffelle_formalized_2021} formally define a \emph{Trace} in the context of forensic science as well as several other concepts. For this paper, only the following concepts are relevant (some definitions are shortened for brevity):

An event is ``a complete collection of related things that have happened (or are happening) in a World within a specific closed interval of time.'' The authors do not make use of this term to represent the output of a typical timestamp extraction from digital data. This then leads to a \emph{Trace} being defined as ``the full modification of the Scene [...] resulting from the Event $E$, completed or not, and subsequent intrinsic events.'' 
An important part here is that the Trace is every modification that happens. However in reality, what we end up observing are \emph{facets of the Trace}. This is explained further as ``when scientists study a Trace [...] only certain facets are observed, and other facets remain unobserved due to lacking knowledge, methods, technology, or resources''. They go on to describe that explanations often do not make that distinction and that a trace is ``very often described according to a particular observed facet and the perspective that is chosen to observe this facet.'' 

Given the focus on timelines in this article, these definitions and this explanation are essential. 
We acknowledge that an event creates many modifications. These consist of different types of digital and non-digital traces. 
Timeline analysis involves the extraction of specific facets that are centred around a timestamp, but also with associated data attached to that timestamp. For example, usually recorded is a timestamp, 
its context (e.g., that it is in the `last modified' field of a Standard Information Attribute (SIA) within an MFT record), 
and other information such as the MFT record ID, the filename that the timestamp relates to, etc. Depending on the software used to generate the timeline, there may be even more data.

Note, while the timestamp from a timeline perspective is associated with a specific object (a specific NTFS file in this example), the timestamp also has a \emph{source}, which in this case is the SIA within the MFT. We argue that considering the source of the timestamp is important when studying tampering.

\section{Event reconstruction and its challenges}
Given that we need to observe one or more facets of a Trace in an attempt to perform event reconstruction, several general problems can frustrate this:

\begin{enumerate}
    \item The passing of time;
    \item The tampering of facets of traces;
    \item The contamination of traces; and
    \item Insufficient knowledge of event traces
\end{enumerate}

The first point refers to the influence of time passing upon the facets of a trace remaining in the environment. Since an investigation starts after the event of interest occurred, it implies that a certain period passed between the time the event took place and the `fixation of the crime scene'. During this interval, the natural course of happenings (intrinsic events) impacts our ability to observe the facets resulting from the earlier event \cite{jaquet-chiffelle_formalized_2021} (\citet{gruber_contamination_2023} refer to this phenomenon as \emph{evidence dynamics}).
Intrinsic events may directly affect facets that can be observed, e.g., rolling logs overwriting facets; or indirectly, e.g., by affecting the environment in which they reside such as clocks drifting over time \cite{stevens_unification_2004}.

Tampering affects the facets of traces that were caused by events or other related events. To hide activities, an attacker or software (e.g., malware) may directly tamper with facets of traces, e.g., manipulating file timestamps or erasing files; or indirectly tamper with them by targeting the entire system (environment),  e.g., by changing the system clock or attacking the Network Time Protocol (NTP). Note that tampering may also occur during the reconstruction process if the chain of custody is not correctly maintained, but this will not be our focus here. 
\citet{palmbach_artifacts_2020} discussed various sources that can be found on the Windows operating system and concluded that none of the sources are reliable as they can be modified or deleted. Consequently, an (advanced) attacker may tamper with facets of traces which may lead to incomplete or out-of-order timelines. Eventually, this will impact the hypothesis creation of the investigator and in the worst case may lead to misinterpretations. 

Other subsequent events may also produce modifications of the environment, e.g., additional facets, \emph{after} the initiation of the investigation process. \citet{gruber_contamination_2023} refer to this phenomenon as the contamination of digital evidence (facets of traces in the newer terminology) and define it as being ``any inadvertent transfer of traits to an object of relevance at any point of the forensic process''. An example of contamination, as given by the authors, may be failing to boot into a forensic live distribution at the acquisition stage, which would affect many timestamps.

Finally, insufficient knowledge of traces from events is a challenge in multiple ways. Firstly, it is necessary to have experimental data to be able to generate a hypothesis as to what a pattern of timeline entries may suggest. There is then the problem that previous knowledge of traces generated by an event may now be incorrect on a new operating system, or even a new version of an application. For example, well-known references such as the SANS poster on NTFS time rules\footnote{\url{https://www.sans.org/posters/windows-forensic-analysis/}} may be outdated and may not consider application-specific deviations \cite{galhuber_time_2021}. This can to some extent be mitigated through specific experiments related to the event that is trying to be inferred, but this is resource-intensive. However, additional limitations in knowledge will reduce the investigator's ability to generate viable alternative hypotheses that would produce the same pattern of observed facets, and that cannot be so easily addressed via experiments.

Overall, there are several problems to consider during the event reconstruction process, particularly when evaluating facets of traces in light of a certain number of hypotheses related to the event.  While these problems may be well understood by some digital forensics practitioners and academics, their impact on the interpretation/reconstruction process is rarely discussed in published research. 

\section{Towards assessing the tamper resistance of sources}
\label{tampering}


In addressing the tampering problem, research studies focused on the aspect of detection, i.e., through which means tampering of sources of facets may be detected by an investigator. Examples would be studies by \cite{oh2024forensic} or \cite{jang2016understanding} who concentrated on detecting timestamp manipulations.

\citet{palmbach_artifacts_2020} present an extended range of sources of facets that may be used to detect tampering of timestamps on Windows: the \$USNjrnl, Prefetch and LNK files, Windows event logs, and the \$LogFile. 
While these sources may contain substantial information, they may also be manipulated or deleted. In addition, these sources may be limited (e.g., fixed size) or unavailable on the system under investigation. 

In such situations, more niche techniques like \emph{digital stratigraphy, time anchors} or \emph{hyper timelines} may be used to detect time inconsistencies, e.g., backdating. Digital stratigraphy, as defined by \citet{casey_digital_2018}, is a method that takes advantage of the knowledge of file systems and the functioning of their allocation algorithms. By analysing the logical arrangement of files on a disk (e.g., as demonstrated in \cite{schneider2024applying}), investigators can infer hypothetical events, provided they understand how the file system allocates and organizes those files. For instance, with the \emph{next fit} strategy, a file created before another is usually positioned at a lower logical address on the disk than the latter. Thus, a file incorrectly ordered according to the functioning of the allocation algorithm may indicate that it was backdated \cite{casey_digital_2018, willassen_finding_2008, tse_forensic_2011}. Another concept is time anchors as discussed by \cite{vanini2024clock} where an anchor is an artefact that contains two timestamps -- one originating from the internal clock and one from an external source. If these timestamps do not match, this indicates tampering.
A third approach is the concept of hyper timelines which focus on events without timestamps and focus on implicit timing information \cite{dreier2024beyond}. Inconsistency in implicit and exploit time information may also indicate tampering. For example, text-based log files are typically ordered by timestamp. If a timestamp is altered without the corresponding line being moved to maintain the correct order, it may indicate tampering.

%

Experiments show that the probability of detecting tampering is high, especially when it concerns file metadata. For instance, it may be difficult for an attacker to forge a timestamp without causing subsequent inconsistencies that should be fixed as well \cite{schneider_tampering_2020, schneider_prudent_2022}. 
Similarly, \citet{galhuber_time_2021} highlight that timestamp forgery tools may modify timestamps in a detectable way, e.g., changing the accuracy of timestamps from nanoseconds to seconds. 
Additionally, only one of the tools they evaluated can modify the entire range of file system timestamps on Windows thus reducing the risks of being detected. 
On top of that, \citet{magnet_2020} highlights that the \$FN timestamps are ``only modified by the Windows kernel and will generally go untouched by antiforensic timestomping tools''. This provides another specific example where, during event reconstruction, one timestamp is more difficult to manipulate than another.

Although there has been work on tampering in digital forensic investigations, none specifically relates to the general problem of event reconstruction. In addition, \citet{neale_fool_2023} stresses the need to consider the reliability of facets, specifically in the context of tampering, and highlights that current research did not provide yet effective means to identify tampering. 
This paper aims to bridge that gap and explores the factors that could be used to formally evaluate the resistance to tampering.


\section{Factors}
\label{sec:factors}
From an examiner's perspective, knowing about the resilience of sources of facets is essential when forming the hypothesis. Our results outline several factors that should be considered when assessing the resistance of sources. As this is a time-consuming process, it likely cannot be done for all sources but should at least be done for the most critical ones. The outcome of this process is a confidence. 

\subsection{Factors to consider in evaluating tamper resistance}
We consider that each source of an observed facet (in our case sources of timeline entries) has a series of properties or factors that affect how resistant it is to active tampering. This is a combination of its intrinsic nature, e.g., the operating system or settings, in which the source resides. 
After reviewing sources of facets used for event reconstruction, seven common factors have been identified which are presented in the upcoming subsections. Each subsection includes a description, examples, and concludes with several categories within that particular factor. This is not exhaustive but has provided useful insight for the case study examples in Sec.~\ref{sec:case_study}. Future work may further refine these factors as more edge-case examples are identified. 

The use of these factors is a general concept. However, as will be seen later, in their current form as intentionally simple tables, they lend themselves to trying to improve the quality of results in mainstream investigations that do not have highly advanced attackers such as \emph{Advanced Persistent Threats (APTs)}. For these more challenging investigations, the factors presented are still relevant, but more resolution in the scoring would be needed, e.g., around advanced command line usage, and perhaps multiple iterations of evaluating tamper resistance. As an example, to score the presence of software that could be used for tampering, scoring is needed for the artefacts used to infer the presence of software. 

There are seven factors identified that can be considered when evaluating the resistance of a source to active tampering: Viability of source to user; permissions; software to edit on system; observed facets of source access; encryption; and file format of the source; organization of the source. 


\subsubsection{Visibility of the source to user}
This factor considers whether the source containing the digital forensic facet is visible to the user. For example, within Windows, a file on the user’s desktop (subject to custom hidden flags being set) is visible to a user. However, the file \texttt{desktop.ini} in the same location is not, which while not highly relevant for forensic investigations illustrates the example. Through a change within the Windows Explorer settings, this file can be made visible\footnote{ `Show hidden files, folders and drives', and `Hide protected operating system files'}. 
Therefore, two other categories of visibility are needed: `Visible via user setting change (enabled)' and `Visible via user setting change (not enabled)'.
There is also the possibility of viewing files via alternative mechanisms such as the terminal (\texttt{ls -a}) or Powershell (\texttt{ls -Hidden}) which also needs to be captured. 

There are additional challenges in evaluating this particular factor. For example, if we consider the Google Chrome Preferences file, which is normally located in the user's home directory\footnote{In \texttt{.../AppData/Local/Google/Chrome/User Data/Default}}. The folder AppData is by default not visible to a user, but the contents are visible (and also editable) via the Google Chrome interface, and therefore we should consider this trace to be user visible. However, the newer Firefox cache, again in a similar location\footnote{In \texttt{.../AppData/Local/Mozilla/Firefox/Profiles/xxx.default- release}}, is not visible by default. While Firefox itself does offer an interface to the cache via the UI (\texttt{about:cache} in the address bar), it provides no mechanism to edit the data. As a result, it should not be considered user visible. 


This leads us to several \emph{user visibility categories} into which a source could fall:
\begin{table}[!ht]
\small
\centering
\begin{tabular}{p{8.3cm}}
    \hline
    User visible via GUI\\
    User visible via other UI method (e.g., terminal)\\
    Visible via user setting change (enabled)\\
    Visible via user setting change (not enabled)\\
    Cannot be made visible\\
    \hline
\end{tabular}
\end{table}


\subsubsection{Permissions}
Another factor to consider is whether permission exists to modify the source. In many systems, some users may have access to everything. Therefore, this consideration should be specific to a particular user and re-evaluated for different users.
For example, on a Windows system, various operations are protected via User Account Control (UAC). The privileges to modify one source over another could be different. Again, taking the Chrome Preferences example, a user can access it without triggering the UAC interface. 
However, regarding `protected' files, on many system configurations the user is an Administrator on the system in question. Therefore, the UAC interface cannot be considered a barrier to accessing protected files as it only requires clicking \emph{Allow}. On domain systems, however, the user of a system may not be an administrator of the system and therefore, in the context of that user, manipulation of a privileged protected source should be considered as protected, e.g., access to another user's Chrome preferences file. 

In specific, more advanced investigations involving intrusions on a system, another consideration is whether there are observed facets of privilege escalation (see Sec. \ref{sec:observed_facets_of_access}). Suppose facets are found suggesting the use of an exploit to elevate a standard user’s privileges. In that case, files previously considered user inaccessible should be considered as if they are user accessible. 

If we consider Plaso parsers, we see a difficulty in applying the permissions factor generically. For example, for the PE parser for portable executables, some will be in user accessible locations, while others will not. Therefore, many sources will need to be evaluated on a source-by-source basis. However, for some that are in one protected location, e.g., prefetch files, the data extracted could be considered to all have the same property on a particular system.

This leads to the following \emph{permission categories}:
\begin{table}[!ht]
\small
\centering
\begin{tabular}{p{8.3cm}}
    \hline
    User accessible\\
    User accessible with prompt\\
    User accessible with password / biometrics\\
    User inaccessible, but observed facets of privilege escalation\\
    User inaccessible\\
    \hline
\end{tabular}
\end{table}

\subsubsection{Software to edit on system}\label{sec:factor_software}
This factor addresses the ease by which a manipulation can be made. As direct manipulation of the physical representation of binary data is not possible, at some level of abstraction, a tool will be used to facilitate manipulations. Initially, this was considered from the perspective of whether a tool was available at all that would allow manipulation. However, this becomes very time-sensitive, e.g., a tampering tool release requires an update of all evaluations and is also difficult to provide a comprehensive answer for all possible file types. Therefore, the most sensible approach was to provide a context-sensitive evaluation of this, i.e., whether there is software on the system in question capable of modifying the trace, or observed facets of the use of such software. 

For example, on a Windows system, RegEdit is available, and therefore, ignoring all other factors in this section such as permissions, an adversary can use it to modify keys and values in the Registry. In contrast, a SQLite database cannot reasonably be modified on a system without third-party software or Powershell library (and installation of either could leave traces). 

Another software option is the presence of a hex editor where all files should be considered as editable. In this case, the relative complexity of making such a modification will be captured via other factors as discussed later. 

There are also some edge cases. For example for Windows event logs, the tool Event Viewer is built into Windows. However, this tool only provides the ability to read event logs and does not provide edit capabilities (other than clearing logs). A default tool for editing event logs on Windows should therefore be considered in the category of `Not on the system'. 
There are also edge cases regarding the Windows Registry. While, as discussed above, Regedit could be used to change data in keys and values, it does not provide access to the last modified timestamp of a key often used in event reconstruction. Therefore, whether a tool is considered available or not, is dependent on the specific facet and what it is being used for. This is by design and will become clear in the examples in Sec.~\ref{sec:case_study}. 

A final example of the situation being important would be tampering with a value in the MountedDevices key. Here, Regedit can be used to access that registry key, but the drive letter values are the REG\_BINARY type, which Regedit will not provide an easy `user interface' method of tampering, but will provide a view that allows the hex to be manually edited. Therefore for this particular trace a summary of `Tool available by default for low-level (hex) editing' is the most appropriate.


In summary, the following \emph{editing software categories} exist:
\begin{table}[!ht]
\small
\centering
\begin{tabular}{p{8.3cm}}
    \hline
    Tool available by default for UI-based editing*\\
    Tool added to this system for UI-based editing*\\
    Tool available by default for low-level (hex) editing\\
    Tool added to this system for low-level (hex) editing\\
    Not on the system\\
    \hline
\end{tabular}
\end{table}

*UI is used rather than GUI as manipulation tools may be a command line


\subsubsection{Observed facets of access}
\label{sec:observed_facets_of_access}
In addition to considering if there is software on a system capable of accessing a source, it is also important to determine if there are observed facets of actual access to that source. For SQLite database viewers, the recent files list associated with the program may provide evidence of a specific database being accessed. In some cases, this information might be even more detailed; for instance, the Registry key \texttt{/NTUSER.DAT/Software/Microsoft/Windows/CurrentVe rsion/Applets/RegEdit} could indicate that a specific key was accessed using RegEdit. However, there may also be scenarios where no explicit traces of a relevant source being accessed are observable, but evidence shows that the associated program was executed.
We define the following \emph{observed facets of access categories}:

\begin{table}[!ht]
\small
\centering
\begin{tabular}{p{8.3cm}}
    \hline
    Observed facets of edit-capable software accessing the specific source\\
    Observed facets of edit-capable software accessing the source\\
    Observed facets of edit-capable software being run\\
    No observed facets of source access\\
    \hline
\end{tabular}
\end{table}


\subsubsection{Encryption}
Another consideration is if the source in question is encrypted. This could be argued as simply an enforcement of permissions, but there are some situations where this is not the same. For example, consider the messaging app Signal, paired with a Windows desktop computer. Here, the files are stored within a user's home folder so they could have access to them. However, the database that stores the data is encrypted using SQLCipher4 \cite{bilz_2021}. The key is available on the system in the \texttt{config.json} file so the database would still be accessible via an SQLbrowser supporting that encryption method. This could be classed as `encrypted, but key recoverable is possible from local system'.

Another complexity within the encryption attribute is the different categories of encryption software implementation. In \citet{hargreaves2009assessing} the difference between file-based, file system-based, container-based and full disk encryption is described. An encrypted single file or an encrypted container, where it is not known whether a password is available to the attacker, should be initially considered as one of the last threee 'Encrypted' variants shown below depending on the specific nature of the key/password storage.
In contrast, the decryption of file-based (e.g. EFS) or full disk (e.g. Bitlocker) when the system is running means that all data is accessible to the system and the encryption may not be relevant for the accessibility of the data contained therein (subject to other permissions and exact abstraction layer of the tampering attempt).

This leads to the following \emph{encryption categories}:

\begin{table}[!ht]
\small
\centering
\begin{tabular}{p{8.3cm}}
    \hline
    No encryption\\
    Encrypted but accessible live (e.g., EFS)\\
    Encrypted (trivial to break) e.g., ROT13 in Windows Registry\\
    Encrypted (key recovery possible from local system)\\
    Encrypted (key stored off device available to user) \\
    Encrypted (key stored off device not available to user)\\
    \hline
\end{tabular}
\end{table}

\subsubsection{(File) Format}
The format of a source also impacts its resilience. 
There are many different (file) formats, and it is impossible to try and list them all here. However, given that in Sec.~\ref{sec:factor_software} we consider hex editors as software that allows editing of a source, it is important to capture the complexity of making such manual edits. 
We identified some broad categories: Sources could be plain text which would be easy to modify, they could be a structured but still text-based format such as XML or JSON, or they could be a binary format, which may be proprietary, proprietary but reverse engineered, or an open format. We also include a `NA' category here, to be used when a source is considered where a user interface tool is available since the low-level format becomes irrelevant at that point. 

The \emph{(file) format categories} can be summarized as follows:

\begin{table}[!ht]
\small
\centering
\begin{tabular}{p{8.3cm}}
    \hline
    Binary proprietary (currently unknown)\\
    Binary proprietary but reverse-engineered (e.g., MFT)\\
    Binary open format (e.g., SQLite)\\
    Text-based machine format (e.g., XML, JSON)\\
    Plain text\\
    NA (GUI edit tool available)\\
    \hline
\end{tabular}
\end{table}


\subsubsection{Organization of the source}
The organization of data within a source (structured, semi-structured, or unstructured) is another factor impacting its resilience. Generally, more organized structures allow easier automation of manipulations (which then allows mass manipulation).
As examples, structured data can often be accessed with tools and a potential manipulation is scriptable. For instance, it is possible to develop a Python script that scans for JPG files and manipulates the EXIF information in the header. In contrast, removing a watermark within an image\footnote{Note that this kind of source is currently not considered by timeline generation software.} requires utilising artificial intelligence (more processing power) or manual work. Consequently, this factor considers indirectly how difficult it is to automate the manipulation of the contents of a source. Categories would therefore include:

\begin{table}[!ht]
\small
\centering
\begin{tabular}{p{8.3cm}}
    \hline
    Structured - timestamp within a known data structure, e.g., MFT\\
    Semi-structured – a timestamp that is stored as a field within JSON but as a text string, e.g., ``Wed 25th Jan 2022 11:35 am''\\
    Unstructured - within a Word document, within the content itself the author has made reference to a date and time of an event\\
    \hline
\end{tabular}
\end{table}



\subsection{Scoring}
The previous section has suggested seven factors, all of which have been argued to affect the extent to which data could be tampered with. Each factor has several options or categories that allow the properties of a specific source or facet within a source to be evaluated and qualitatively described. That level of granularity is sometimes required, for example, one key within a Windows registry may have different properties to another, e.g., the user autorun keys vs the contents of keys in the SAM file, the latter being inaccessible via Regedit to even admin users on a live machine \cite{laiho_2020}.

Let us now consider an event reconstruction that relies on some observed facets. If we can assign scores to each of the categories within each of the factors used to describe the source of a facet, then we can use this to begin to evaluate the reliability of an event reconstruction from a tamper resistance perspective. 
At present, there is no meaningful absolute score that could be assigned to those categories, nor data on the relative importance of each category. However, in other areas, quantitative measures are used which are broader values and are used to rate one situation over another. This is used as inspiration here. For the scoring, we decided to borrow concepts from security risk assessment \cite{ross2012guide} where the determination of risk can be seen as a function of harm and the likelihood of its occurrence. As the harm is difficult to predict, we use it to express the \emph{tampering concern} of the source from that factor's perspective (the tampering concern is the inverse of tampering resistance).
Given a source (e.g., Windows Registry) and a factor (e.g., software to edit), we define three degrees of severity:
\begin{description}
    \item[high (3)] means that there is the highest tampering concern of the source from that factor's perspective 
    \item[moderate (2)]  means that there is a moderate tampering concern.
    \item[low (1)] means that there is a low tampering concern. 
\end{description}

We then looked at each factor for each category and assigned a severity where a higher number means that manipulating a source is easier. For instance, the category `Cannot be made visible' in the `user visibility' factor received a low score (1). Consequently, if a source is assigned this category, the tampering concern is low / the tampering resistance is high.
Even though some categories have received the same severity, we keep the qualitative descriptions separate to facilitate a more granular analysis in the future and to provide provenance as to why a source has been given a particular rating.

An important note is that each factor is independent of one other so a 3 in user visibility is not equivalent to a 3 in permissions. This means that at this stage a meaningful computed sum is not possible, but it does mean that sources with particular weaknesses can be easily identified. It is crucial to emphasize that evaluating the tamper resistance of the sources used for event reconstruction, as discussed above, is just as important as the numerical scores themselves. This consideration should carry significant weight when contributing to a C-Score assessment.

\section{Case Study examples}\label{sec:case_study}
We can now consider some examples taking into account the factors and scores. For the production of these examples, a template spreadsheet has been created that captures the factors discussed earlier and the available options are in a dropdown menu, and from that, a color-coded score is displayed. The template is available\footnote{The template and examples are available via Google Sheets. To use it, open the link and create a copy (file $>$ duplicate): \url{https://docs.google.com/spreadsheets/d/1DnfYMtp-rmzp3dGt9SxRo2Jb83ruZHdRMStFz3PzZQ8/}} and can be used to review an event reconstruction. The tables are used in these examples and could be used by an investigator to assist in structuring an assessment of tamper resistance of sources. We also consider them to be a step towards a more automated analysis in the future where some fields could be automatically populated. 

To perform a review of an event reconstruction, the expected facets that result from an event need to be known. This can be achieved either through existing documented forensic research, via live logging tools such as Procmon, or using timeline generation software such as Plaso. The relevant observed facets need to be identified and the source from which they are extracted determined. 
Then the `tampering concern' of the source can be reviewed and classified according to the factors discussed above. 

A simple example is provided on file creation on NTFS (see Sec.~\ref{sec:cs_file_creation}), and an extended example with multiple variations is given on USB device connection on Windows (see Sec.~\ref{sec:cs_usb_device}). 
 Note, that the goal of this section is not to explore exhaustively
each factor category but to improve the understanding of
their usability in evaluating the tamper resistance of sources.  



\begin{table}[ht!]
    \scriptsize
    \begin{tabular}{|c|p{1.8cm}|p{4.5cm}|c|}
    \hline
    n&     Factors    & Category  & Score   \\\hline
    \multicolumn{4}{|c|}{\cellcolor[HTML]{F2F3F4}SI attribute}                                                                                                                    \\ \hline
    1&     User visible               & Cannot be made visible           &\cellcolor[HTML]{ABEBC6}1   \\
    2&     Permissions                & User inaccessible          &\cellcolor[HTML]{ABEBC6}1  \\
    3&     Software to edit           & Tool added to this system for UI-based editing           &\cellcolor[HTML]{F5B7B1}3   \\
    4&     Facets of access           & Observed facets of edit-capable software being run           &\cellcolor[HTML]{F9E79F}2  \\
    5&     Encryption                 & No encryption          &\cellcolor[HTML]{F5B7B1}3  \\
    6&     File format                & NA (UI edit tool available)          &\cellcolor[HTML]{F5B7B1}3   \\
    7&     Structural                 & Structured          &\cellcolor[HTML]{F9E79F}2  \\\hline
\multicolumn{4}{|c|}{\cellcolor[HTML]{F2F3F4}FN attribute}  \\ \hline
    8&     User visible               & Cannot be made visible           &\cellcolor[HTML]{ABEBC6}1   \\
    9&     Permissions                & User inaccessible          &\cellcolor[HTML]{ABEBC6}1  \\
    10&     Software to edit           & Not on the system           &\cellcolor[HTML]{ABEBC6}1   \\
    11&    Facets of access           & No observed facets of source access           &\cellcolor[HTML]{ABEBC6}1  \\
    12&     Encryption                 & No encryption          &\cellcolor[HTML]{F5B7B1}3  \\
    13&     File format                & Binary proprietary but reverse engineered         &\cellcolor[HTML]{F9E79F}2   \\
    14&     Structural                 & Structured          &\cellcolor[HTML]{F9E79F}2  \\\hline
    \end{tabular}
    \caption{Computed severity for two creation timestamps within the MFT given the presence of timestomp.\label{fig:timestomp}}
\end{table}



\subsection{File Creation reconstruction when timestomp is present}\label{sec:cs_file_creation}
As a first simple example, to illustrate the process, we consider event reconstruction of the `creation of a file' on Windows but use the example from \citet{magnet_2020} referenced earlier regarding timestomp. Looking at just the two creation timestamps within an MFT entry, we create  Table~\ref{fig:timestomp} which consists of two sub-tables, one for each facet (the MFT SI and FN attributes). 
On rows 3 and 10 one can see the two different scores assigned because timestomp is capable of modifying the SIA attribute but not the FN attribute. This then cascades into the file format information which is not relevant when a tool is available in SIA, but is relevant for the FN. We conclude from this that in this context, the timestamp in the FN attribute is more tamper resistant than the SIA attribute, which aligns with the intuitive findings in \cite{magnet_2020} . An important highlight here is that the granularity of `source' in this case needs to be at the resolution of MFT attributes, one for the SIA, and one for the FN attribute, since they have different properties.


\subsection{USB Device Connection}\label{sec:cs_usb_device}
For the second example, we consider the connection of a USB device on Windows as summarized in Table~\ref{fig:usb_device}. There are several known locations where modifications are made (setupapi.dev.log, Windows Registry, Event Logs).

\begin{table}[ht!]
    \scriptsize
    \begin{tabular}{|c|p{1.8cm}|p{4.5cm}|c|}
    \hline
     n    & Factors    & Category  & Score  \\\hline
\multicolumn{4}{|c|}{\cellcolor[HTML]{F2F3F4}Windows/INF/setupapi.dev.log}                                                                                                                    \\ \hline
    1&     User visible               & User visible via GUI           &\cellcolor[HTML]{F5B7B1}3   \\
    2&     Permissions                & User accessible          &\cellcolor[HTML]{F5B7B1}3  \\
    3&     Software to edit           & Tool available by default for UI-based editing           &\cellcolor[HTML]{F5B7B1}3   \\
    4&     Facets of access           & Observed facets of edit-capable software being run           &\cellcolor[HTML]{F9E79F}2  \\
    5&     Encryption                 & No encryption          &\cellcolor[HTML]{F5B7B1}3  \\
    6&     File format                & Plain text          &\cellcolor[HTML]{F5B7B1}3   \\
    7&     Structural                 & Structured          &\cellcolor[HTML]{F9E79F}2  \\\hline
\multicolumn{4}{|c|}{\cellcolor[HTML]{F2F3F4}System/ControlSetxxx/Enum/ USBSTOR/}  \\ \hline
    8&     User visible               & User visible via GUI           &\cellcolor[HTML]{F5B7B1}3   \\
    9&     Permissions                & User accessible with prompt         &\cellcolor[HTML]{F5B7B1}3  \\
    10&     Software to edit           & Tool available by default for UI-based editing           &\cellcolor[HTML]{F5B7B1}3  \\
    11&     Facets of access           & No observed facets of source access           &\cellcolor[HTML]{ABEBC6}1  \\
    12&     Encryption                 & No encryption          &\cellcolor[HTML]{F5B7B1}3  \\
    13&     File format                & NA (UI edit tool available)         &\cellcolor[HTML]{F5B7B1}3   \\
    14&     Structural                 & Structured          &\cellcolor[HTML]{F9E79F}2  \\\hline
\multicolumn{4}{|c|}{\cellcolor[HTML]{F2F3F4}System/MountedDevice}  \\ \hline
    15&     User visible               & User visible via GUI          &\cellcolor[HTML]{F5B7B1}3   \\
    16&     Permissions                & User accessible with prompt          &\cellcolor[HTML]{F5B7B1}3 \\
    17&     Software to edit           & Tool available by default for UI-based editing           &\cellcolor[HTML]{F5B7B1}3   \\
    18&     Facets of access           & No observed facets of source access           &\cellcolor[HTML]{ABEBC6}1  \\
    19&     Encryption                 & No encryption          &\cellcolor[HTML]{F5B7B1}3  \\
    20&     File format                & Binary proprietary but reverse engineered       &\cellcolor[HTML]{F9E79F}2   \\
    21&     Structural                 & Structured          &\cellcolor[HTML]{F9E79F}2  \\\hline
\multicolumn{4}{|c|}{\cellcolor[HTML]{F2F3F4}Event Logs}  \\ \hline
    22&     User visible               & User visible via GUI           &\cellcolor[HTML]{F5B7B1}3   \\
    23&     Permissions                & User accessible          &\cellcolor[HTML]{F5B7B1}3  \\
    24&     Software to edit           & Not on the system           &\cellcolor[HTML]{ABEBC6}1   \\
    25&     Facets of access           & No observed facets of source access           &\cellcolor[HTML]{ABEBC6}1  \\
    26&     Encryption                 & No encryption          &\cellcolor[HTML]{F5B7B1}3  \\
    27&     File format                & Binary proprietary but reverse engineered         &\cellcolor[HTML]{F9E79F}2   \\
    28&     Structural                 & Semi-structured          &\cellcolor[HTML]{F9E79F}2  \\\hline
    \end{tabular}
    \caption{Computed severity for four sources used for event reconstruction of USB device connection.\label{fig:usb_device}}
\end{table}

One can see that given some specific conditions on this particular system:  notepad has been run but no reference to setupapi.dev.log (row 4), and there are no observed facets of Regedit running on this system (rows 11, 18)). As a direct comparison between the sources, the Windows event logs should be considered the most difficult to tamper with from the set. Thus, if there were conflicting timestamps, from a tampering perspective only, and in the absence of other information, the times in the event logs should be considered the most reliable of the set. 

An extension of this would be if older copies of the Windows Registry were available within Shadow Copies as illustrated in Table~\ref{fig:reg_shadow_copy}. These can be accessed via the command line mounting of restore points (vssadmin) \cite{hargreaves2008windows}. For that data, the tamper resistance score changes since they are not directly accessible without mounting shadow copies and Regedit cannot directly access those versions of the registry.

\begin{table}[ht!]
    \scriptsize
    \begin{tabular}{|c|p{1.8cm}|p{4.5cm}|c|}
    \hline
    n    & Factors    & Category  & Score   \\\hline
    \multicolumn{4}{|c|}{\cellcolor[HTML]{F2F3F4}Registry within Shadow Copy}                                                                                                                    \\ \hline
    1&     User visible               & Visible via user setting change (not enabled)           &\cellcolor[HTML]{ABEBC6}1   \\
    2&     Permissions                & User accessible with prompt          &\cellcolor[HTML]{F5B7B1}3  \\
    3&     Software to edit           & Not on the system           &\cellcolor[HTML]{ABEBC6}1    \\
    4&     Facets of access           & No observed facets of source access           &\cellcolor[HTML]{ABEBC6}1  \\
    5&     Encryption                 & No encryption          &\cellcolor[HTML]{F5B7B1}3  \\
    6&     File format                & Binary proprietary but reverse engineered          &\cellcolor[HTML]{F5B7B1}3   \\
    7&     Structural                 & Structured          &\cellcolor[HTML]{F9E79F}2  \\\hline
    \end{tabular}
    \caption{Computed severity of a Registry within a Windows Shadow Copy\label{fig:reg_shadow_copy}}
\end{table}

In addition, Table.~\ref{fig:usb_no_admin} shows that the difference in tamper resistance is significantly different when a corporate system is considered and the end-user accessibility of many of the sources is reduced compared with Table.~\ref{fig:usb_device}.



\subsection{Discussion}
These examples illustrate the difference in the tamper resistance of different sources that are frequently used for event reconstruction. The factors that have been identified have been argued to have an effect on tamper resistance. While there may be additional factors or categories within those factors, this can easily be accommodated. In particular, new categories can be trivially added and appropriate scores assigned depending on if it is easier or harder to tamper with a source with those properties. 

\begin{table}[ht!]
    \scriptsize
    \begin{tabular}{|c|p{1.8cm}|p{4.5cm}|c|}
    \hline
     n&    Factors    & Category  & Score   \\\hline
    \multicolumn{4}{|c|}{\cellcolor[HTML]{F2F3F4}Windows/INF/setupapi.dev.log}                                                                                                                    \\ \hline
    1&     User visible               & Cannot be made visible           &\cellcolor[HTML]{ABEBC6}1  \\
    2&     Permissions                & User inaccessible          &\cellcolor[HTML]{ABEBC6}1  \\
    3&     Software to edit           & Not on the system          &\cellcolor[HTML]{ABEBC6}1   \\
    4&     Facets of access           & No observed facets of source access           &\cellcolor[HTML]{ABEBC6}1  \\
    5&     Encryption                 & No encryption          &\cellcolor[HTML]{F5B7B1}3  \\
    6&     File format                & Plain text          &\cellcolor[HTML]{F5B7B1}3   \\
    7&     Structural                 & Semi-structured          &\cellcolor[HTML]{F9E79F}2  \\\hline
\multicolumn{4}{|c|}{\cellcolor[HTML]{F2F3F4}System/ControlSetxxx/Enum/ USBSTOR/}  \\ \hline
    8&     User visible               & Cannot be made visible           &\cellcolor[HTML]{ABEBC6}1  \\
    9&     Permissions                & User inaccessible         &\cellcolor[HTML]{ABEBC6}1 \\
    10&     Software to edit           & Not on the system          &\cellcolor[HTML]{ABEBC6}1\\
    11&    Facets of access           & No observed facets of source access           &\cellcolor[HTML]{ABEBC6}1  \\
    12&     Encryption                 & No encryption          &\cellcolor[HTML]{F5B7B1}3  \\
    13&     File format                & Binary proprietary but reverse engineered         &\cellcolor[HTML]{F9E79F}2   \\
    14&     Structural                 & Structured          &\cellcolor[HTML]{F9E79F}2  \\\hline
\multicolumn{4}{|c|}{\cellcolor[HTML]{F2F3F4}Event Logs}  \\ \hline
    15&     User visible               & Cannot be made visible           &\cellcolor[HTML]{ABEBC6}1  \\
    16&     Permissions                & User inaccessible         &\cellcolor[HTML]{ABEBC6}1 \\
    17&     Software to edit           & Not on the system          &\cellcolor[HTML]{ABEBC6}1\\
    18&     Facets of access           & No observed facets of source access           &\cellcolor[HTML]{ABEBC6}1  \\
    19&     Encryption                 & No encryption          &\cellcolor[HTML]{F5B7B1}3  \\
    20&     File format                & Binary proprietary but reverse engineered         &\cellcolor[HTML]{F9E79F}2   \\
    21&     Structural                 & Semi-structured          &\cellcolor[HTML]{F9E79F}2  \\\hline
    \end{tabular}
    \caption{Computed severity for three sources when a corporate system is considered.\label{fig:usb_no_admin}}
\end{table}

The examples demonstrated the use for some Windows event reconstructions, but the proposed categories are platform-independent. The examples did not consider external sources of facets, e.g., network server logs, although we acknowledge their importance. For brevity, this paper focuses on single evidence items which allows clarity in terms of the intrinsic resilience of individual sets of local traces. 

It has also been shown how factors can be risk scored to highlight differences between sources that are used for event reconstruction. A simplistic high, moderate, and low system has been chosen to provide an easy evaluation of sources, which given that this is currently necessarily a manual process seems an appropriate level of granularity.

At present manual scoring of sources is necessary, and many are situation/environment dependent, e.g., is the user admin or not. Some could be standardized, for example, if the machine is Windows and the user is the only account and therefore admin, then the Registry is always accessible (clicking through the UAC prompt) and Regedit will always be available. This offers the potential for prepopulating some of the scores. At present, we only offer a manual process, which is more time-consuming but does encourage deeper thought about the nature of the sources being used for event reconstruction. It is also necessary to have a good knowledge of system behaviour to understand what a user can and cannot access on a system, plus details such as the binary nature of sources. However, we argue that a requirement of detailed knowledge of digital systems and their behaviour should not be an onerous requirement to improve the reliability of event reconstruction. 

It should also be stressed that due to the detailed nature of this process, this could be performed only for specific examples where the event reconstruction that being performed is critical to a case and offers an opportunity to further improve reliability and provides some defence to questions about whether tampering could have occurred; e.g., that an assessment was performed according to a structured framework. This is most value when used in a targetted manner e.g., when there are conflicting timestamps identified in an event reconstruction. However, it could also be undertaken less regularly using a `dip sample' method as part of a general quality assurance assessment of processes. 

In the future, with a programmatic assessment of these factors affecting reliability, the manual overhead does become negligible (potentially with the source knowledge preprogrammed), and also a more detailed quantitative measure could be used as a replacement for the simplistic high/moderate/low options. 
The proposed score can assist investigators in reconstructing events when multiple sources are not available or in cases where they differ. It can be used to express the tamper resistance of specific sources or used conjointly with the C-Scale \cite{CASEY2020200888} to help assess the tamper resistance of sources when expressing the strength of observed facets in light of competing hypotheses (e.g., understanding and evaluating indicators such as in C-value C2 ``\textit{only one source of evidence that is not difficult to tamper with}''). 
Whilst the scoring system is intended to remain partly subjective (to retain some insights into an investigator's reasoning process), we imagine it being implemented in software such as Plaso to give an indicator of the tamper resistance of sources in timelines (if prepopulating the scores is feasible, which will part of future work). Finally, the scoring system has educational applications for raising awareness of the risks of relying on observed facets for event reconstruction and encouraging critical questioning of their reliability.

\section{Conclusions and further work}
Event reconstruction is a critical part of the digital forensic process. We introduced a process for reviewing and scoring sources contributing to timeline-based event reconstruction. Our analysis revealed that some commonly relied-upon facets, such as USB device connections, may not be as resistant to tampering as often assumed. While this does not preclude their use, it underscores the importance of understanding their limitations.

The primary takeaway from this work is the importance of considering tamper resistance when reconstructing events. Although our proposed factors and scoring system represent an initial framework, future iterations and refinements will enhance its applicability, especially as edge cases emerge. The general principle of evaluating the tamper resistance of traces will be invaluable for improving the reliability of event reconstruction, handling uncertainty, and reducing errors in the process.

Finally, while some of these identified factors will be obvious to seasoned investigators, and many will understand the reliability issues associated with certain sources, there is a clear need within what is now referred to as digital forensic science, to formalize definitions and make explicit that which is currently tacit. Furthermore, providing categories within these influencing factors, and including concrete examples of how they can be used provides the foundation for more formal and potentially future quantitative evaluation of the trustworthiness or indeed reliability of reconstructed events in a digital forensic investigation.


\bibliographystyle{elsarticle-num-names} 
\bibliography{refs.bib}


%





\end{document}